\documentclass[prd,twocolumn,showpacs,floats,letterpaper,nofootinbib]{revtex4}
\usepackage{amssymb,amsmath,amsfonts}
\usepackage{color}
\usepackage{subfigure}

\usepackage{graphicx}

\newcommand{\bea}{\begin{eqnarray}}
\newcommand{\eea}{\end{eqnarray}}
\newcommand{\bean}{\begin{eqnarray*}}
\newcommand{\eean}{\end{eqnarray*}}

\newcommand{\PRL}{Phys. Rev. Lett.}
\newcommand{\PRD}{Phys. Rev. D}

\begin{document}

\title{Impact of Point Source Clustering on Cosmological Parameters
  with CMB Anisotropies} \author{Paolo Serra$^1$, Asantha Cooray$^1$,
  Alexandre Amblard$^1$, Luca Pagano$^2$, and Alessandro
  Melchiorri$^2$} \affiliation{$^1$Center for Cosmology, Department of
  Physics and Astronomy, University of California, Irvine, CA
  92697.\\ $^2$Dipartimento di Fisica ``G. Marconi'' and INFN, sezione
  di Roma, Universita' di Roma ``La Sapienza'', Ple Aldo Moro 5,
  00185, Roma, Italy.  }

\begin{abstract}
The faint radio point sources that are unresolved in cosmic microwave
background (CMB) anisotropy maps are likely to be a biased tracer of
the large-scale structure dark matter distribution.  While the
shot-noise contribution to the angular power spectrum of unresolved
radio point sources is included either when optimally constructing the
CMB angular power spectrum, as with WMAP data, or when extracting
cosmological parameters, we suggest that clustering part of the point
source power spectrum should also be included.  This is especially
necessary at high frequencies above 150 GHz, where the clustering of far-IR 
sources is expected to dominate the shot-noise level of the angular power spectrum at tens of
arcminute angular scales of both radio and sub-mm sources.  We make an estimate of source clustering of
unresolved radio sources in both WMAP and ACBAR, and marginalize over
the amplitude of source clustering in each CMB data set when model
fitting for cosmological parameters.  For the combination of WMAP
5-year data and ACBAR, we find that the spectral index changes from
the value of $0.963 \pm 0.014$ to $0.959 \pm 0.014$ (at $68 \%$ c.l.)
when the clustering power spectrum of point sources is included in
model fits.  While we find that the differences are marginal with and
without source clustering in current data, it may be necessary to
account for source clustering with future datasets such as Planck,
especially to properly model fit anisotropies at arcminute angular
scales. If clustering is not accounted and point sources are modeled with a shot-noise only out to $l \sim 2000$,
the spectral index will be biased by about 1.5$\sigma$.
\end{abstract}

\pacs{98.70.Vc,98.65.Dx,95.85.Sz,98.80.Cq,98.80.Es}

\maketitle

\section{Introduction} 

As discussed in a variety of papers, unresolved radio point sources
are an important foreground in temperature anisotropy maps of the
cosmic microwave background (CMB) \cite{zotti,tof1, wright, tegmark}.
With Wilkinson Microwave Anisotropy Probe (WMAP) data \cite{bennett},
the difference in the CMB power spectra determined at various
frequency channels and the cross power spectra between the channels,
has allowed the unresolved point source contamination to be
constrained with an amplitude $A_{\rm ps}=(0.011 \pm 0.001)
\mu$K$^2$-sr \cite{nolta} for the foreground power spectrum, when
scaled to the Q-band.  In the WMAP analysis, this point source
amplitude is taken to be a constant in $C_l$, similar to the case of a
shot-noise type power spectrum for unresolved point sources. This
shot-noise, with an appropriate scaling in frequency, is then removed
from each of the power spectra when constructing the final WMAP
temperature anisotropy power spectrum \cite{hinshaw}.

While the WMAP estimate on the point source correction is consistent
with a point source power spectrum dominated by the shot-noise, this
estimate is dominated by measurements relative to the Q-band
\cite{nolta}.  Since the WMAP temperature anisotropy power spectrum is
based on V- and W-band data \cite{dunkley}, and the point source
correction has a larger uncertainty in V- and W- bands \cite{nolta},
it is unclear if a simple shot-noise correction fully describes point
sources in the WMAP temperature anisotropy power spectrum.  To account
for uncertainty in the amplitude of point-source shot-noise in
parameter estimates, the WMAP likelihood contains an additional
marginalization of the uncertainty of $A_{\rm ps}$, but the best-fit
amplitude of point-sources remain fixed to the a priori determined
value \cite{hinshaw}.  We also note that alternative approaches have
been considered to estimate the amplitude of point-sources
\cite{Huffenberger1,Huffenberger2}, though these works also
concentrated on establishing the shot-noise correction.

Beyond the shot-noise, unresolved radio point sources are likely to
have a clustered distribution on the sky as they are expected
to be a biased tracer of the large-scale structure. Thus, the angular
power spectrum of sources contains not just a shot-noise but also a
clustering piece determined by the dark matter power spectrum, point
source bias, and the redshift distribution. Existing calculations
suggest that the shot-noise part of the power spectrum from bright,
rare sources dominates clustering at low radio frequencies, especially
when the flux threshold for point source removal is at the level of
$\sim$ 1 Jy \cite{GonzalezNuevo:2004fj,tof1}.  Thus, the assumption of
a shot-noise point source contribution to the angular power spectrum
of temperature anisotropies is likely to be adequate for low-frequency
bands of WMAP such as the Q-band, but may not be appropriate at high
frequencies, such as the W band, whose data are used in the
temperature power spectrum.

The approach using a shot-noise spectrum with an uncertainty that is
marginalized over in the WMAP likelihood is is bit different from the
approach advocated by the WMAP team to account for another foreground
in CMB data involving the Sunyaev-Zel'dovich (SZ) effect from galaxy
clusters. There, the angular power spectrum is estimated based on a
model for the cluster distribution and gas properties \cite{Komatsu},
with an overall uncertainty in the amplitude of the SZ power spectrum
captured by a free parameter which is then freely varied as a nuisance
parameter when best-fit cosmological parameter values and their
uncertainties using a Markov-Chain Monte-Carlo (MCMC) code
\cite{lewis,spergel,dunkley}.

Since the clustering component of the angular power spectrum of
unresolved radio sources may be important, it could be that simply
including point-sources as a shot-noise correction in the V/W-band
WMAP temperature anisotropy power spectrum results in biased estimates
of cosmological parameters, especially for parameters like the
spectral index of density perturbations, which has been discussed
previously in the context of uncertainties related to point-source
shot-noise amplitude \cite{Huffenberger1}.  Moreover, current CMB
analyses make use of the combination of datasets such as WMAP and
ACBAR which have different treatments related to how point sources are
accounted in the parameter fits. At the ACBAR frequency of 150 GHz
\cite{reichardt}, the clustering of sources may need to be
included properly, especially given that the high angular resolution
of ACBAR also allows removal of sources down to a lower flux density
level, where the shot-noise associated with rare, bright sources may
be subdominant. The clustering of point sources could also account for
some fraction of the excess arcminute-scale anisotropies detected by
ACBAR \cite{tof2,Cooray3}.

Given the lack of adequate details related to the exact clustering
power spectrum of radio sources in datasets such as WMAP and ACBAR, we
make a general estimate of the angular power spectrum of radio
point-sources and include the overall amplitude of clustering as an
additional nuisance parameter to be included and marginalized over
when estimating for cosmological parameters. For this, we make use of
number counts at 95 GHz \cite{sadler} and assume the redshift
distribution of high-frequency radio sources follows the same
distribution as estimated for NVSS \cite{condon} sources at low
frequencies \cite{Ho}. As some of these assumptions are likely to be
invalid to some extent, we do not fix the clustering spectrum to our
model but allow the overall amplitude to vary and marginalize over
that uncertainty when constraining cosmological parameter. Thus, the
uncertainty in our predictions related to clustering of sources is
unlikely to dominate and we confirm this by noting that the
differences to best-fit parameter values with point source clustering
included are not significant, especially for the case of combined WMAP
and ACBAR data.  Moreover, our estimate of clustering is consistent
with the allowed level of point source correction in V- and W-band
combination of WMAP, as measured in terms of differences in the power
spectra \cite{nolta}.

While the differences in estimated cosmological parameters are small,
a proper estimate of cosmological parameters with clustering included
is useful for a proper statistical analysis on important scientific
results such as on the extent to which the spectral index of density
perturbations $n_s$ is different from the Harrison-Zel'dovich value at
-1.  While the impact on current datasets is small, for future data
such as Planck that probe down to smaller scales over a wide range of
frequencies, we suggest that it will be necessary to account for
clustering of point sources when model fitting cosmological
parameters.

This paper is organized as follows: we first discuss the angular
clustering power spectrum of radio points sources.  Section~III
discusses model fits to recent CMB anisotropy data from WMAP and
ACBAR. We discuss our results and conclude with a summary in
Section~IV. 

\section{Clustering of Radio Point Sources}

The angular power spectrum of radio sources, in units of $(\mu K)^2$,
generally contains two components \cite{Scott,Oh:2003sa}
\begin{equation}
C_l = \left(\frac{\partial B_\nu}{\partial T}\right)^{-2}\left[\int_0^{S_{\rm cut}} S^2 \frac{dN}{ds} \, dS  + \bar{I}^2 w_l \right]\, ,
\end{equation}
where $w_l$ is the Legendre transform of the angular correlation
function $w(\theta)$ of unresolved radio point sources, $\bar{I}$ is
the average intensity (in flux units) produced by these sources
\begin{equation}
\bar{I} = \int_0^{S_{\rm cut}} S \frac{dN}{dS} \, dS \, ,
\end{equation}
 and the conversion factor from flux to antenna temperature using the
 CMB black-body spectrum, $B_\nu(T=2.726{\rm K})$, is
\begin{equation}
\frac{\partial B_\nu}{\partial T}=\frac{2k_B}{c^2} \left(\frac{k_BT}{h}\right)^2 \frac{x^4 e^x}{(e^x-1)^2} \, ,
\end{equation}
where $x\equiv h\nu/k_BT=\nu/56.84$ GHz is the dimensionless
frequency. This conversion can be simplified as $\partial
B_\nu/\partial T= [(99.27 {\rm Jy} sr^{-1})/\mu K] x^4 e^x/(e^x-1)^2$.

Since we know little about the clustering of radio sources below the
point source detection limit W- and V- bands of WMAP and at 150 GHz of
ACBAR, we make use of a simplified set of assumptions to estimate the
source clustering.  In general, the angular power spectrum of the
source sources can be written with the halo model \cite{Cooray2} such
that $w_l$ is
\begin{equation}
w_l^{\rm lin} = \int dz\, \frac{dr}{dz} a^2(z) \frac{n^2(z)}{d_A^2(z)} P_{ss}\left(k=\frac{l}{d_A},z\right) \, ,
\end{equation}
where $P_{ss}(k,z)$ is the three-dimensional power spectrum of radio
sources as a function of redshift.  In the halo model, source
clustering at large angular scales can be described with the linear
matter power spectrum scaled by a constant and scale-free bias factor:
\begin{equation}
P_{ss}(k) \approx b^2_s P^{\rm lin}(k) \, ,
\end{equation}
where the source bias factor, when combined with an estimate of the
number density of sources, provide some information on the halo mass
scale associated with those sources through the luminosity- or
flux-averaged halo occupation number $\langle N(M,z)\rangle$, halo
bias $b_{\rm halo}(M,z)$, and the halo mass function $dn/dM$
\cite{Cooray2}:
\begin{eqnarray}
b_s = \frac{1}{\bar{n}_g}\int dM\, \frac{dn}{dM}(z)\, b_{\rm halo}(M,z) \langle N(M,z) \rangle\, .
\end{eqnarray}
At small angular scales, clustering traces the non-linear power
spectrum generated by the so-called 1-halo term.  Separating the
occupation number to central and satellite radio sources, $\langle
N(M)\rangle= \langle N_{\rm s} \rangle + \langle N_{\rm c} \rangle$,
the 1-halo power spectrum is
\begin{eqnarray}
&& P^{1h}(k) = \\\
&& \int dM\; n(M)\; \frac{2 \langle N_{\rm s} \rangle \langle N_{\rm c} \rangle u(k|M)  + \langle N_{\rm s} \rangle^2 u^2(k|M)}{\bar{n}_g^2} \, . \nonumber
\end{eqnarray}
Here, $u(k|M)$ is the normalized density profile in Fourier space. At
deeply non-linear scales, however, the shot-noise term is expected to
dominate the clustering spectrum, but the transition scale may lie
larger than the shot-nose amplitude.  The above form of the 1-halo
term allows us to easily understand a simple behavior. If radio
sources occupy dark matter halos such that there is only one source
per halo, regardless of the halo mass, then with $N_{\rm s}=0$,
$P^{1h}=0$. Thus, the 1-halo term only exists to the extent that more
than one radio source occupies a halo. While there is limited
information on the halo occupation properties of radio sources at the
frequencies of interest, observations at frequencies around 30 GHz
suggest that multiple radio sources are found in large dark matter
halos such as groups and clusters, though at 30 GHz, the central
galaxy tends to be the dominant bright source in most galaxy clusters
\cite{Cooray4}.

Given the lack of detailed knowledge on the clustering of radio
sources or even ingredients such as luminosity functions or exact
redshift distributions that can be used to generate a reliable halo
model for the radio source population using approaches such as the
conditional-luminosity functions that are used to describe clustering
of optical or IR and far-IR galaxies \cite{Amblard}, we make several
approximations. First we note that at large angular scales, $C_l
\approx \bar{I}^2\langle b_{s}^2\rangle w_l^{\rm lin}$
\cite{Scott}. To calculate the angular power spectrum, we assume that
unresolved sources trace the same large-scale structure as
low-frequency NVSS sources and estimate $\bar{I}$ by integrating over
the number counts at 95 GHz as estimated by \cite{sadler}.  We use 95
GHz as a first estimate here since it is close to both WMAP channels
on one end and ACBAR at the other end.  We make use of the redshift
distribution estimates for NVSS to calculate clustering at high
frequencies \cite{Ho}.  In addition to linear clustering, we also
include a non-linear correction to the angular clustering using a
1-halo model that assumes a simple power-law occupation number for
satellite galaxies with $N_s(M) \sim M^\beta$ with $\beta= 0.85$ when
$M > 10^{12.5}$ M$_{\odot}$. The typical bias factor estimate for
sources from this occupation number is about $1$ at $z \sim 1$. Before
calculating anisotropies for CMB, we verified that our prediction for
source clustering, when applied for low-frequency sources, generally
agrees with measurements from the literature \cite{Blake,Overzier}.

In Fig.~1, we show the angular power spectrum of radio sources as
fluctuations in the CMB temperature $C_l$, and a comparison to the
difference in power spectra of V and W-bands of WMAP \cite{nolta}. For
this comparison, we follow the same procedure as the WMAP analysis
\cite{nolta} and scaled the power spectrum to Q-band (40.7 GHz) and
estimate $A_{\rm ps}=r(Q)^{-2}C_l^{\rm PS}(Q)$ with
$r(\nu)=(e^x-1)^2/(x^2e^x)$ with a numerical value for the Q-band of
$r(Q)=1.089$.  In Fig~1, we have converted  our 
estimate of $C_l$'s from 95 GHz counts to Q-band with average spectral
index of bright resolved WMAP sources with $\alpha \approx
  -0.09$ \cite{wright}, with the scaling $\nu^{\alpha-2}r(\nu)$ for temperature units instead of intensity. 
In addition to clustering we also include the V- and W-band combined estimate of
shot-noise in WMAP data with a value of $0.007$ $\mu$K$^2$-sr, when
scaled to the Q-band.

As shown in Fig.~1, the sum of this shot-noise and the clustering we
estimate is consistent with the allowed amplitude of point source
correction from Nolta et al. \cite{nolta}.  Though there are larger
uncertainties in point source estimates of the V and W-band, we cannot
simply rule out that the unresolved sources only contribute with a
shot-noise type power spectrum. If Q-band is also included, as shown
in Ref.~\cite{nolta}, the differences are more consistent with a
shot-noise power spectrum, as expected for low-frequencies since
Q-band dominates such an estimate.  Since the final WMAP power
spectrum is composed of V- and W-band data, we find that there is some
motivation to include source clustering when estimating cosmological
parameters.

While the clustering amplitude is unconstrained, this is of little
concern to us in the cosmological parameter estimation since we will
not fix the clustering of radio sources to a pre-determined model, but
parameterize the overall amplitude of clustering with a free
parameter.  Thus, when model fitting the data we parameterize the
clustering part as $P_{\rm WMAP}C_l$ and consider $P_{\rm WMAP}$ as a
nuisance parameter that captures all uncertainties in our calculation,
which includes the spectral index of sources from counts at 95 GHz to
WMAP band, their redshift distribution, parameters of the halo model,
among others. When quoting cosmological parameter measurements, we
marginalize the likelihood over $P_{\rm WMAP}$.  This approach is
consistent with how the WMAP team included the effect of SZ angular
power spectrum in parameter estimation with a parameter $A_{\rm SZ}$
that is freely varied.  Note that we only include a model for the
source clustering since the CMB power spectrum released by the WMAP
team already has a shot-noise removed from the data for point sources
when combining V and W-data to a final power spectrum.

In addition to clustering of sources as relevant for WMAP, we also
include clustering of point sources as related to ACBAR data
\cite{reichardt}.  In parameter estimation, unlike the WMAP team that
fitted and removed a constant shot-noise spectrum for unresolved radio
sources when estimating an optimal power spectrum from data, the ACBAR
team included the shot-noise of radio sources as an extra component in
their model fits. Thus, while we only include clustering spectrum of
radio sources for WMAP, for ACBAR data, we include both a clustering
spectrum and a shot-noise for radio sources. The shot-noise was taken
to be consistent with estimates made by the ACBAR team and the
clustering component was taken by simply frequency scaling the same
WMAP spectrum to 150 GHz with the same scaling as the one involved
with the shot-noise part.

\begin{figure}[!t]
\begin{center}
\includegraphics[scale=0.8,angle=-90,width=8cm]{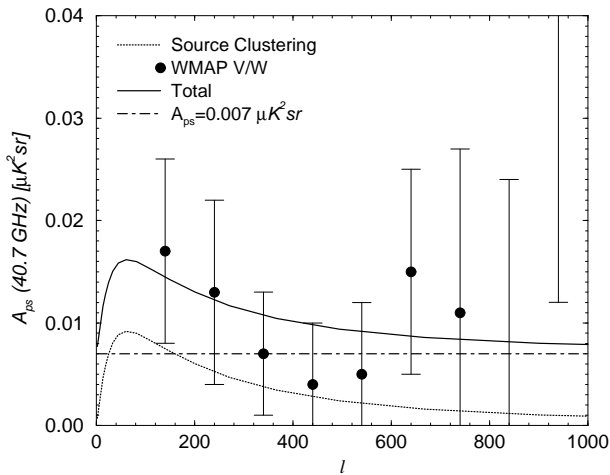}
\end{center}
\caption{The amplitude of point source correction to the V- and W-band
  WMAP angular power spectrum.  The data points show the measurement
  from the WMAP team scaled to the Q-band.  We ignore the corrections
  with Q-band as the final power spectrum from the WMAP team uses only
  V- and W-band data.  The dotted line shows our model for the point
  source clustering (scaled to Q-band) with $P_{\rm WMAP}=1$, while
  the solid line shows the total clustering spectrum arising from
  point sources with clustering and the shot-noise. The shot-noise is
  taken to be the same as estimated by the WMAP team for V and W-band
  data with a value $A_{\rm PS}=0.007$ $\mu$K$^2$-sr }
\end{figure}

\section{Cosmological Parameters with Clustered Point Sources}

The method we use to estimate cosmological parameters is based on the
publicly available Markov Chain Monte Carlo package CosmoMC
\cite{lewis} with a convergence diagnostics based on the Gelman and
Rubin statistic. We used WMAP 5-year data \cite{Komatsu5yr} (both
temperature and temperature-polarization cross-correlation) alone and
in combination with ACBAR data \cite{reichardt}.  We only account for
point sources in temperature anisotropies. Since WMAP polarization
data do not probe small angular scales, where polarized point sources
contribute, ignoring point sources in polarization is a safe
assumption.

In our estimates we make use of the flat $\Lambda$CDM cosmological
model with 6 cosmological parameters: baryon density $\Omega_bh^2$,
dark matter density $\Omega_ch^2$, reionization optical depth $\tau$,
ratio of the sound horizon to the angular diameter distance at the
decoupling measured by $\theta$, amplitude of the curvature
perturbation $A_s$ (with flat prior on $log(A_s)$) and spectral index
$n_s$; these two last parameters are both defined at the pivot scale
$k_0=0.002/$ Mpc as in \cite{dunkley}.  To this set we include $A_{\rm
  SZ}$, the amplitude of SZ contribution, and two parameters $P_{\rm
  WMAP}$ and $P_{\rm ACBAR}$ for the amplitude of point-source
clustering.  To study the impact of point sources on running of the
spectral index and estimates of the tensor-to-scalar ratio, we also
consider additional runs where these quantities are varied.

\subsection{WMAP and ACBAR data}

When estimating parameters with existing WMAP and ACBAR data, with
point sources and SZ included, the total CMB anisotropy spectrum is
\begin{equation}
C_l^{\rm tot} = C_l^{\rm CMB} + C_l^{\rm PS}+C_l^{\rm SZ} \, .
\end{equation}
The point-source angular power spectrum contains two parts as
discussed: $C_l^{\rm PS}=C_l^{\rm sn}+C_l^{\rm c}$, but since the WMAP
team removed the shot-noise when combining data to a single estimate
of the power spectrum, we take $C_l^{\rm WMAP}=C_l^{\rm
    CMB}+C_l^{\rm c}+C_l^{\rm SZ}$. There is a slight complication
here since $C_l^{\rm c}$ is a combination of the clustering in
V- and W-bands, and we make the simple assumption here that the
clustering of sources between these two bands can be scaled by a
constant while the shape remains the same. The uncertainty in the
variation of the point source clustering with frequency, to some
extent, is not expected to be a significant issue since we allow the
overall amplitude to vary with $P_{\rm WMAP}$.  Note that the same
complication exists for $C_l^{\rm SZ}$, but in this case the frequency
dependence is known exactly.

\begin{figure}[!t]
\begin{center}
\includegraphics[scale=0.8,angle=-90,width=8cm]{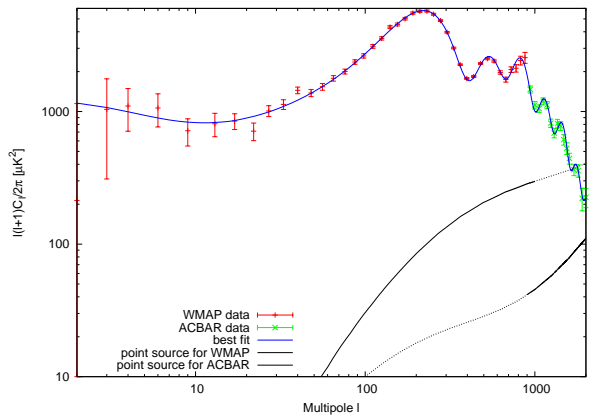}
\end{center}
\caption{Best-fit CMB angular power spectrum for WMAP 5-year data and
  ACBAR with a comparison to measurements. We also show the input
  power spectra of point source clustering for WMAP (middle line) and
  ACBAR data (bottom line) with $P_{\rm WMAP}=P_{\rm ACBAR}=1$ (see
  text for details). In the case of WMAP, we show custering $C_l^c$ part only as the shot-noise is removed from the data, while for ACBAR we
show the total.}
\end{figure}

  \begin{figure*}
  \begin{center}
    \begin{tabular}{ccc}
      \resizebox{50mm}{!}{\includegraphics{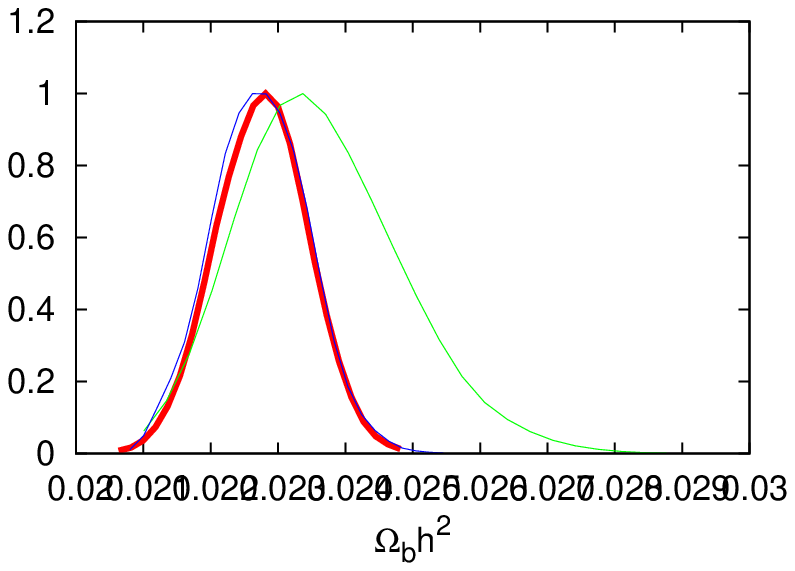}} &
      \resizebox{50mm}{!}{\includegraphics{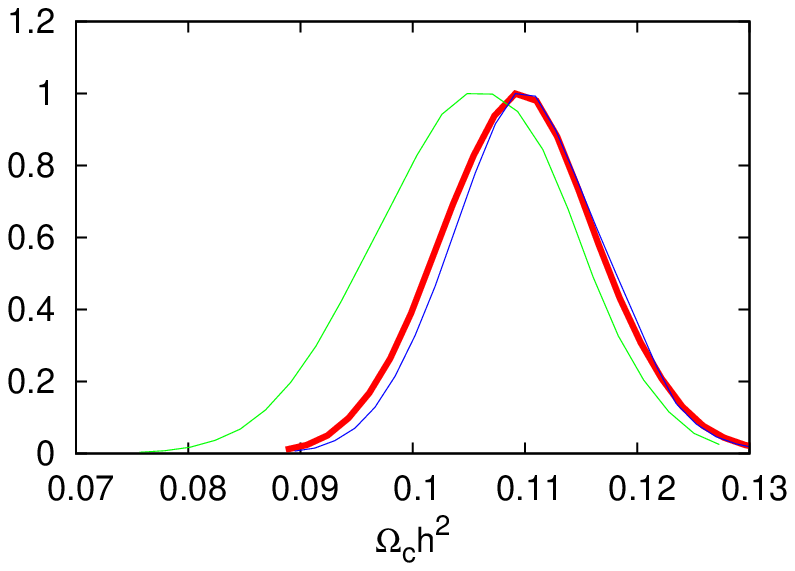}} &
      \resizebox{50mm}{!}{\includegraphics{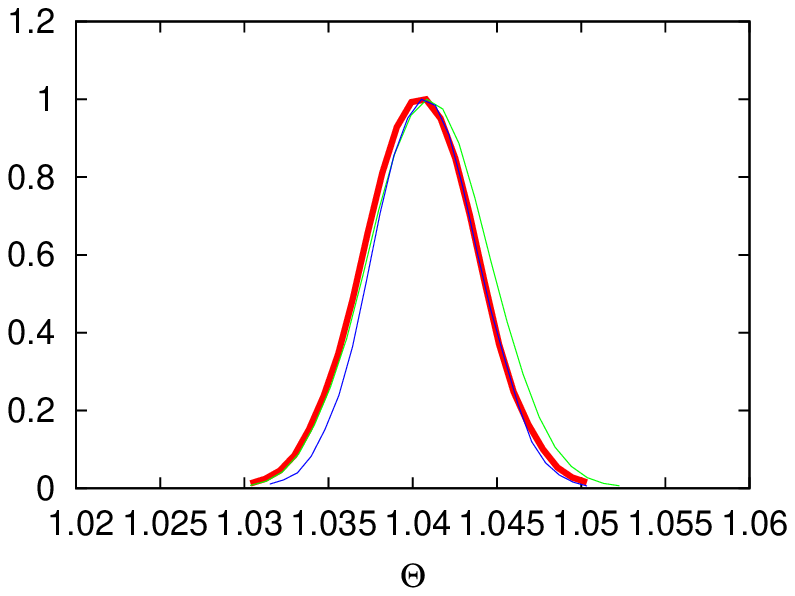}}  \\
      \resizebox{50mm}{!}{\includegraphics{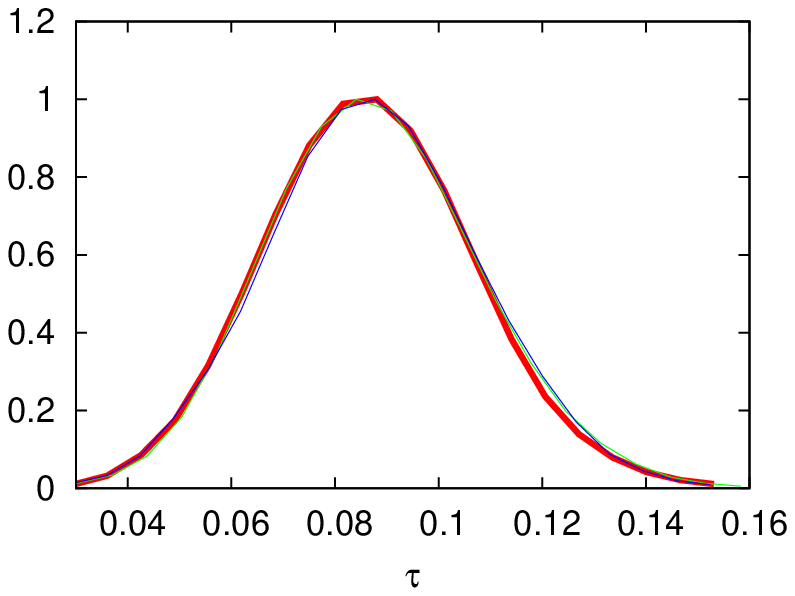}} &
      \resizebox{50mm}{!}{\includegraphics{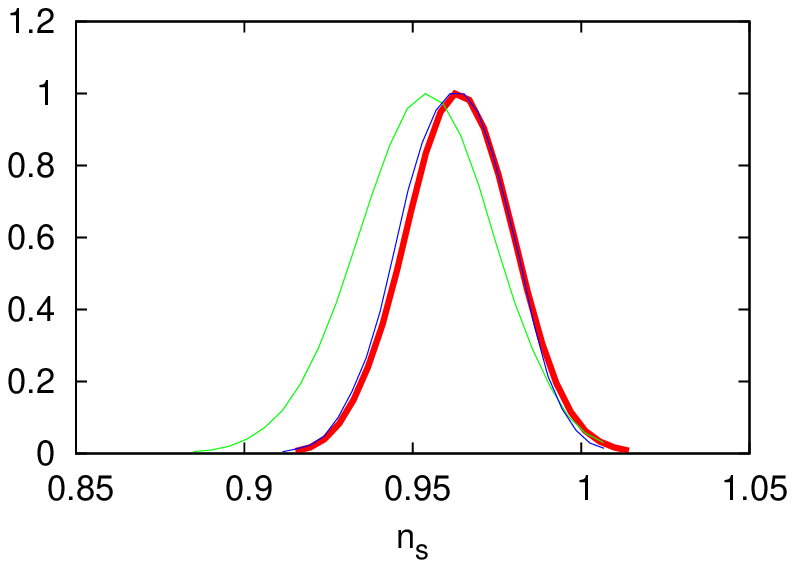}} &
      \resizebox{50mm}{!}{\includegraphics{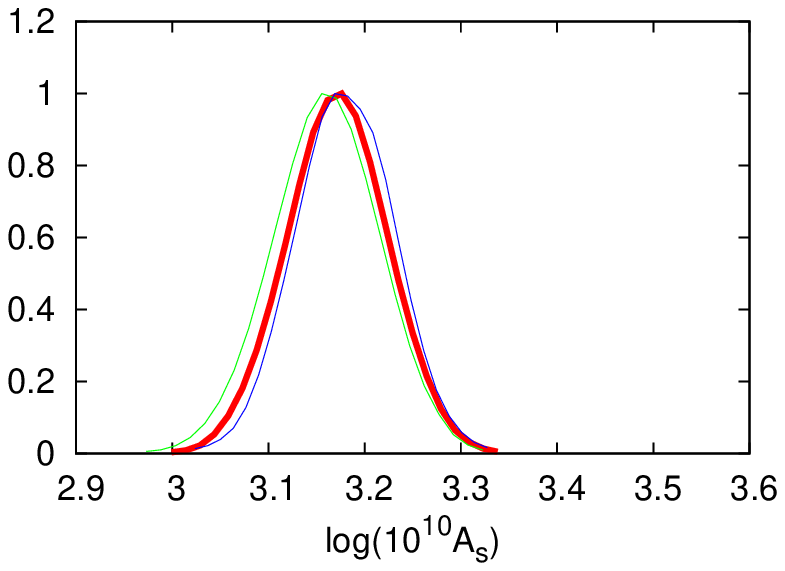}} \\
      \resizebox{50mm}{!}{\includegraphics{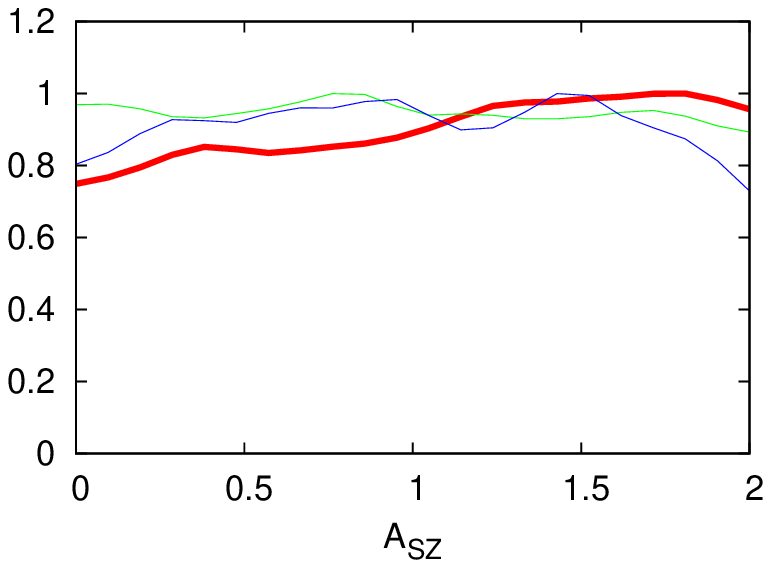}} &
      \resizebox{50mm}{!}{\includegraphics{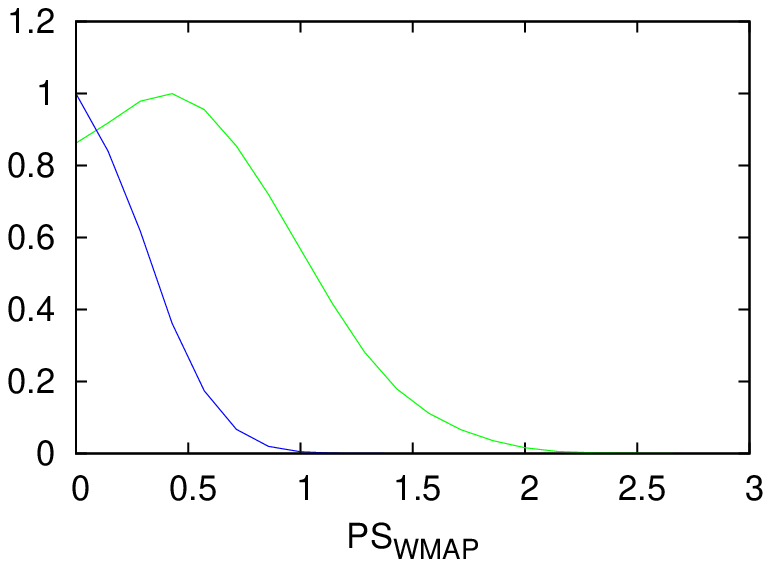}} &
      \resizebox{50mm}{!}{\includegraphics{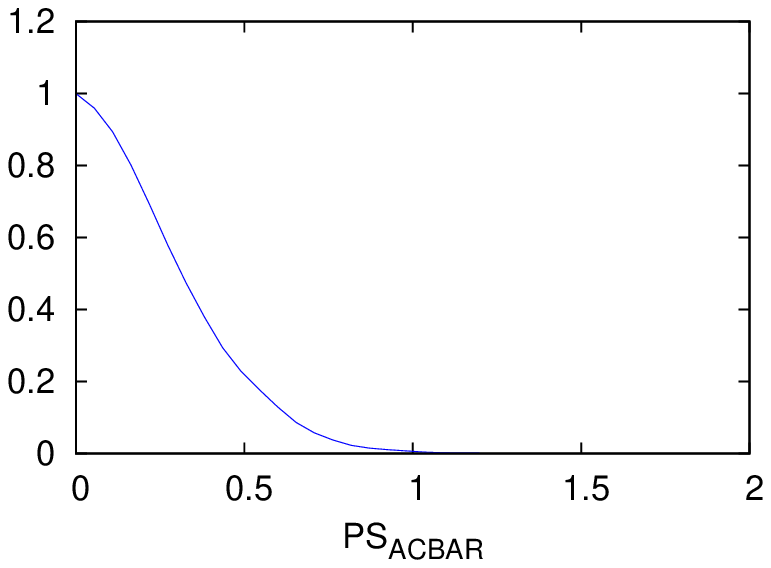}} \\
    \end{tabular}
    \caption{Marginalized parameter constraints for WMAP without
      clustering (red line), WMAP with point source clustering (green
      line), and WMAP+ACBAR with point source clustering signal for
      both (blue line). From left to right, each of the panels show
      the constraint on the baryon density, cold dark matter density,
      the ratio of sound horizon to the angular diameter distance at
      the decoupling (top panels), optical depth, spectral index,
      amplitude of curvature perturbations (middle panels), and SZ
      normalization, WMAP point source normalization, and ACBAR point
      source normalization (lower panels), respectively.  The spectral
      index and the amplitude of perturbations is measured at
      $k=0.002$ Mpc$^{-1}$.}
    \label{test4}
  \end{center}
\end{figure*}

\begin{table*}
\caption{Mean values and marginalized $68 \%$ c.l. limits for several
  cosmological parameters from WMAP and WMAP+ACBAR, with and without
  clustering of point sources (see text for details).}
\begin{center}
\begin{tabular}{|c|c|c|c|c|c|c|}
\hline
Parameter & WMAP 5-yr  & WMAP 5-yr &         WMAP 5-yr  & WMAP+ACBAR & WMAP+ACBAR\\
          &            & with clustering &  with clustering &   &     with clustering \\
          &            & $C_l^{c}\times P_{\rm WMAP}$  & $C_l^{c}\times(0<P_{\rm WMAP} < 1)$&  &$C_l^{c}\times(0<P_{\rm WMAP} < 1)$ \\
          &             &                             &                                     & & $C_l^{c}\times(0<P_{\rm ACBAR} < 1)$  \\
\hline \hline
$\Omega_bh^2$&$0.02277\pm0.00062$&$0.02397_{-0.00104}^{+0.00103}$& $0.02366_{-0.00083}^{+0.00084}$ &$0.02269_{-0.00060}^{+0.00059}$&$0.02298_{-0.00065}^{+0.00063}$\\
$\Omega_{\rm c}h^2$&$0.1093_{-0.0063}^{+0.0064}$&$0.1025_{-0.0075}^{+0.0075}$& $0.1041\pm0.0069$  &$0.1103\pm0.0059$&$0.1092_{-0.0059}^{+0.0056}$\\
$\Omega_{\Lambda}$&$0.744_{-0.029}^{+0.030}$&$0.780\pm0.034$& $0.772_{-0.030}^{+0.031}$ &$0.740_{-0.029}^{+0.028}$&$0.747_{-0.027}^{+0.028}$\\
$n_s$&$0.965\pm0.014$&$0.949_{-0.017}^{+0.018}$& $0.953\pm0.016$&$0.963\pm0.014$&$0.959\pm0.014$\\
$\tau$&$0.087\pm0.017$&$0.088\pm0.017$& $0.088_{-0.017}^{+0.018}$ &$0.087_{-0.017}^{+0.016}$&$0.086_{-0.016}^{+0.018}$\\
$\Delta^2_R$&$(2.39\pm0.10)\cdot10^{-9}$&$(2.33\pm0.10)\cdot10^{-9}$& $(2.35\pm0.10)\cdot10^{-9}$ &$(2.41\pm0.10)\cdot10^{-9}$&$(2.40\pm0.10)\cdot10^{-9}$\\
\hline
$\sigma_8$&$0.793\pm0.036$& $0.726\pm 0.056$& $0.742\pm0.047$ &$0.798\pm0.033$ &$0.784\pm0.033$\\
$\Omega_m$&$0.256_{-0.030}^{+0.029}$&$0.22\pm0.034$& $0.228_{-0.030}^{+0.031}$&$0.260_{-0.028}^{+0.0029}$&$0.253_{-0.028}^{+0.0027}$\\
$H_0$&$72.1_{-2.6}^{+2.7}$&$76.3_{-3.9}^{+4.0}$&$75.3\pm3.4$ &$71.7\pm2.5$&$72.5\pm2.6$\\
$z_{reion}$&$11.0\pm1.4$&$10.5_{-1.3}^{+1.4}$&$10.7\pm1.4$ &$11.0\pm1.4$&$10.8\pm1.4$\\
$t_0$&$13.68_{-0.13}^{+0.14}$&$13.49\pm0.19$& $13.54\pm0.16$&$13.69_{-0.12}^{+0.13}$&$13.65\pm0.13$\\
$A_{SZ}$&$1.04_{-0.69}^{+0.68}$&$1.00\pm0.68$& $1.00\pm0.68$&$0.98_{-0.66}^{+0.67}$&$0.91_{-0.65}^{+0.68}$\\
$P_{\rm WMAP}$&$--$&$<1.38 (2\sigma)$& $<0.93 (2\sigma)$ &$--$&$<0.46 (2\sigma)$\\
$P_{\rm ACBAR}$&$--$&$--$&$--$ &$--$&$<0.56 (2\sigma)$\\
\hline
\end{tabular}
\label{table:1}
\end{center}
\end{table*}

As described, in addition to WMAP 5-year data, we also include ACBAR
data at large multipoles. To avoid complicating the analysis when
different datasets overlap, which requires a calculation of the
covariance matrix between different experiments, as they observe the
same sky, in the likelihood calculation, we take the same approach as
the WMAP team and use WMAP data out to $\ell < 900$ and ACBAR data
from $900 < \ell < 2000$.  For ACBAR data, we make a separate estimate
of the angular power spectrum by scaling the flux-cut of unresolved
point sources to be at the lower flux threshold and in agreement with
previous shot-noise estimates \cite{reichardt}.  Again, we include an
overall uncertainty in the ACBAR angular power spectrum of radio
sources, in this case the sum of clustering and shot-noise terms of
the power spectrum, with $P_{\rm ACBAR}$.  In Fig.~2, we show the
angular power spectrum of CMB anisotropies with best-fit cosmological
model for WMAP and ACBAR data, as well as the two input power spectra
for point source clustering with both $P_{\rm WMAP}=P_{\rm ACBAR}=1$.

\begin{figure*}[]
\centering
\subfigure[]{\includegraphics[width=8cm]{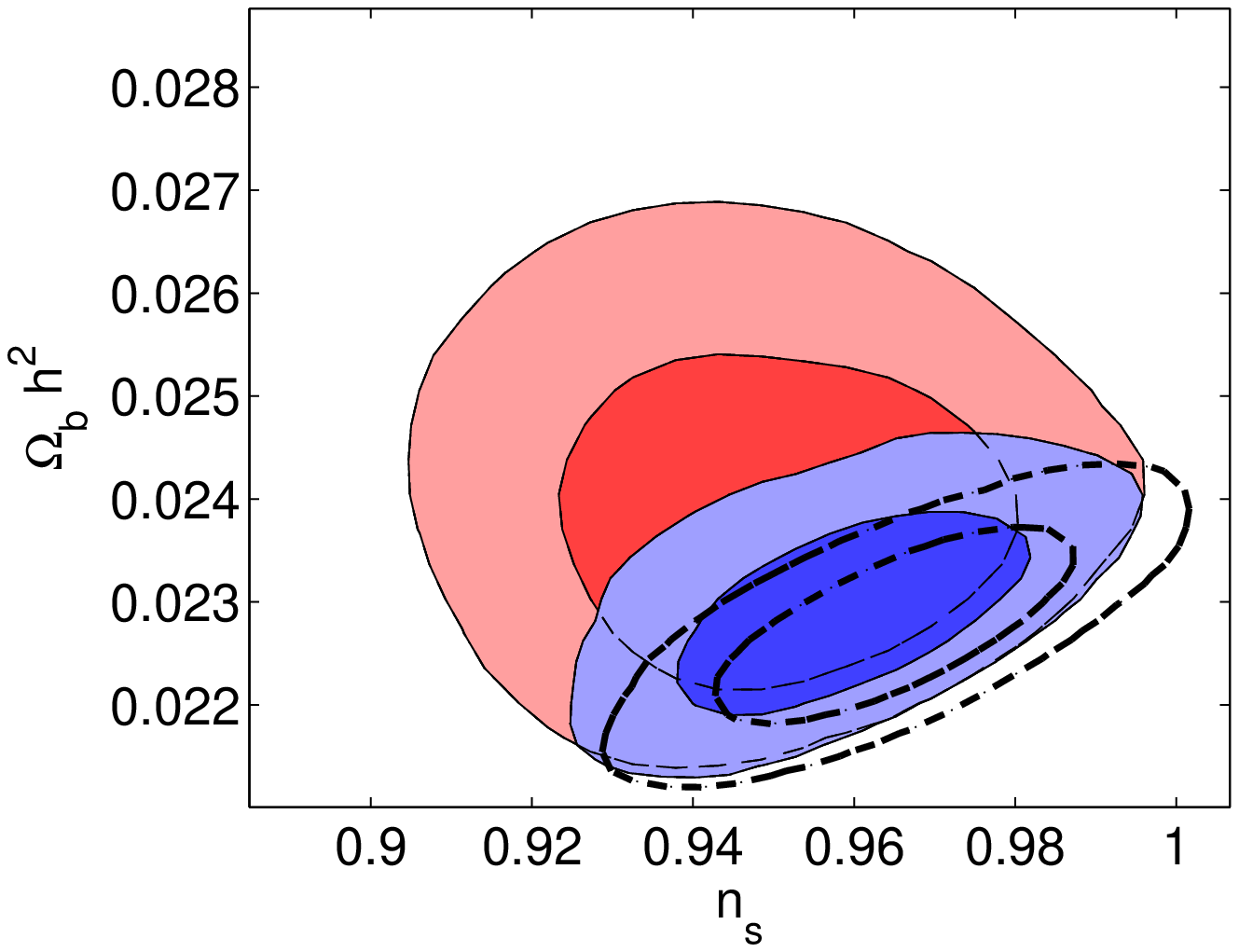}}
\hspace{0.1cm}
\subfigure[]{\includegraphics[width=8cm]{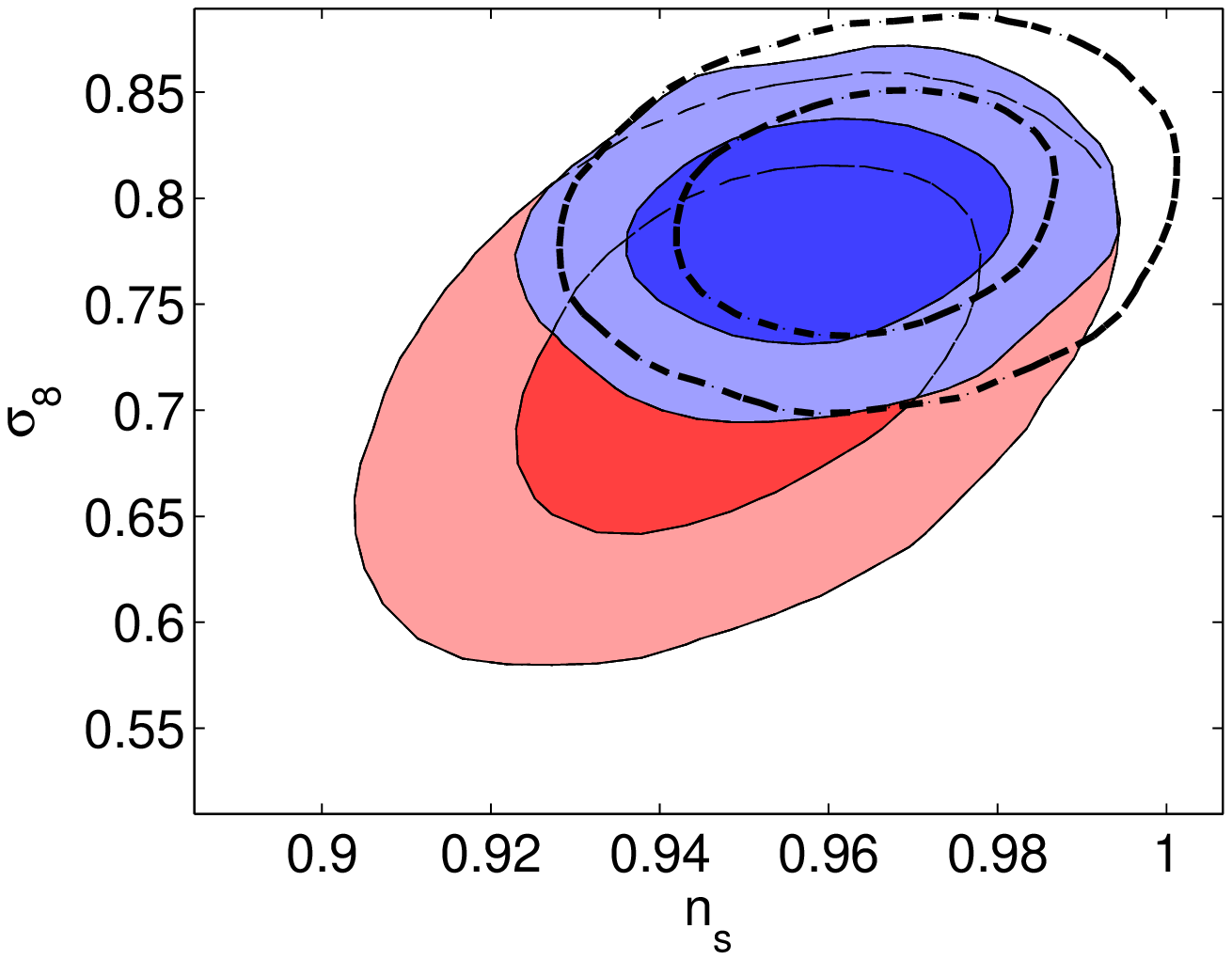}}
\caption{\label{burn-in1} Two-dimensional marginalized distributions
  showing the $68\%$ and $95\%$ confidence level contours for $n_s$
  vs. $\Omega_bh^2$ (left) and $n_s$ vs. $\sigma_8$ (right) with WMAP
  5-years data alone (empty contours), WMAP 5-years data with
  clustered point sources (red contours), and WMAP 5-years data+ACBAR
  data with point source clustering (blue contours).}
\end{figure*}

\begin{table}[!ht]
\caption{Constraints on spectral index $n_s$, running of the spectral
  index ${dn_s/d\ln k}$ and tensor to scalar ratio $r$ from WMAP+ACBAR
  with and without point source signal. These values are evaluated at
  $k=0.002$ Mpc$^{-1}$.}
\begin{center}
\begin{tabular}{cll}
\hline
parameter                    & WMAP+ACBAR & WMAP+ACBAR \\
    &    & with clustering\\
\hline \hline
 n$_s$        &   $1.036\pm0.046$ & $1.041\pm0.050$  \\
 $dn_s/d\ln k$    & $-0.038\pm0.024$  & $-0.044\pm0.025$ \\
\hline
 n$_s$        & $0.981\pm0.019$     & $0.977_{-0.019}^{+0.020}$ \\
 r          & $<0.36 (2\sigma)$   &   $<0.39 (2\sigma)$      \\
\hline
\end{tabular}
\end{center}
\end{table}

Since we only include the clustering term of unresolved point
sources for WMAP data, we follow the WMAP team's approach on the
shot-noise term and marginalize the likelihood over the uncertainty
related to point source shot-noise term A$_{\rm ps}$ using the public
WMAP likelihood routine.  This uncertainty only makes a small
difference in best-fit cosmological parameters \cite{nolta}.  The
results related to $\Lambda$CDM runs are summarized in Table~I.  In
the case where we do not consider clustering of point sources, we
essentially recover the same results as Ref.~\cite{dunkley}, with
small differences at the level of 0.1$\sigma$, which we believe is due
to differences in the numerical codes and the convergence criteria.

With clustering of point sources included, however, the spectral index
estimated with WMAP 5-year data alone changes from $0.965 \pm 0.014$
with point source shot-noise only to $0.949 \pm 0.018$ with source
clustering in addition to the shot-noise from the WMAP likelihood.
This is a difference of about $1\sigma$, but this large difference
primarily comes from the fact that $P_{\rm WMAP}$ is largely
unconstrained by the data with a $2\sigma$ upper limit of 1.38.  The
change in $n_s$ is captured by a similar change in $\sigma_8$ with
values changing from 0.793 $\pm 0.036$ without clustering to $0.726
\pm 0.056$ with clustering.  If we put a prior that $P_{\rm WMAP}$ is
uniform between 0 and 1, $n_s=0.953\pm0.016$ and the difference from
the case with clustering ignored is about $\sim0.8\sigma$.

\begin{figure*}[]
\centering
\subfigure[]{\includegraphics[width=8cm]{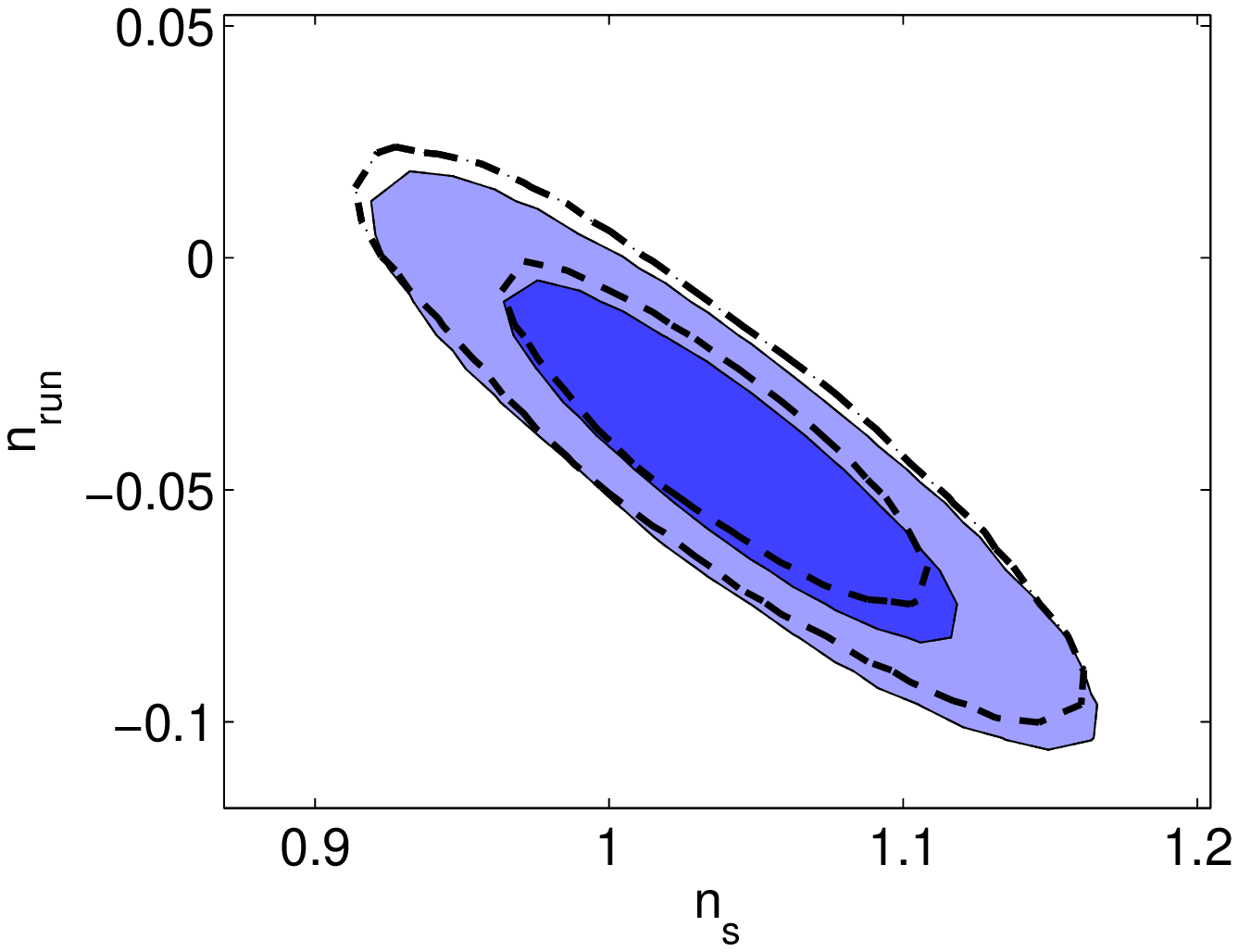}}
\hspace{0.1cm}
\subfigure[]{\includegraphics[width=8cm]{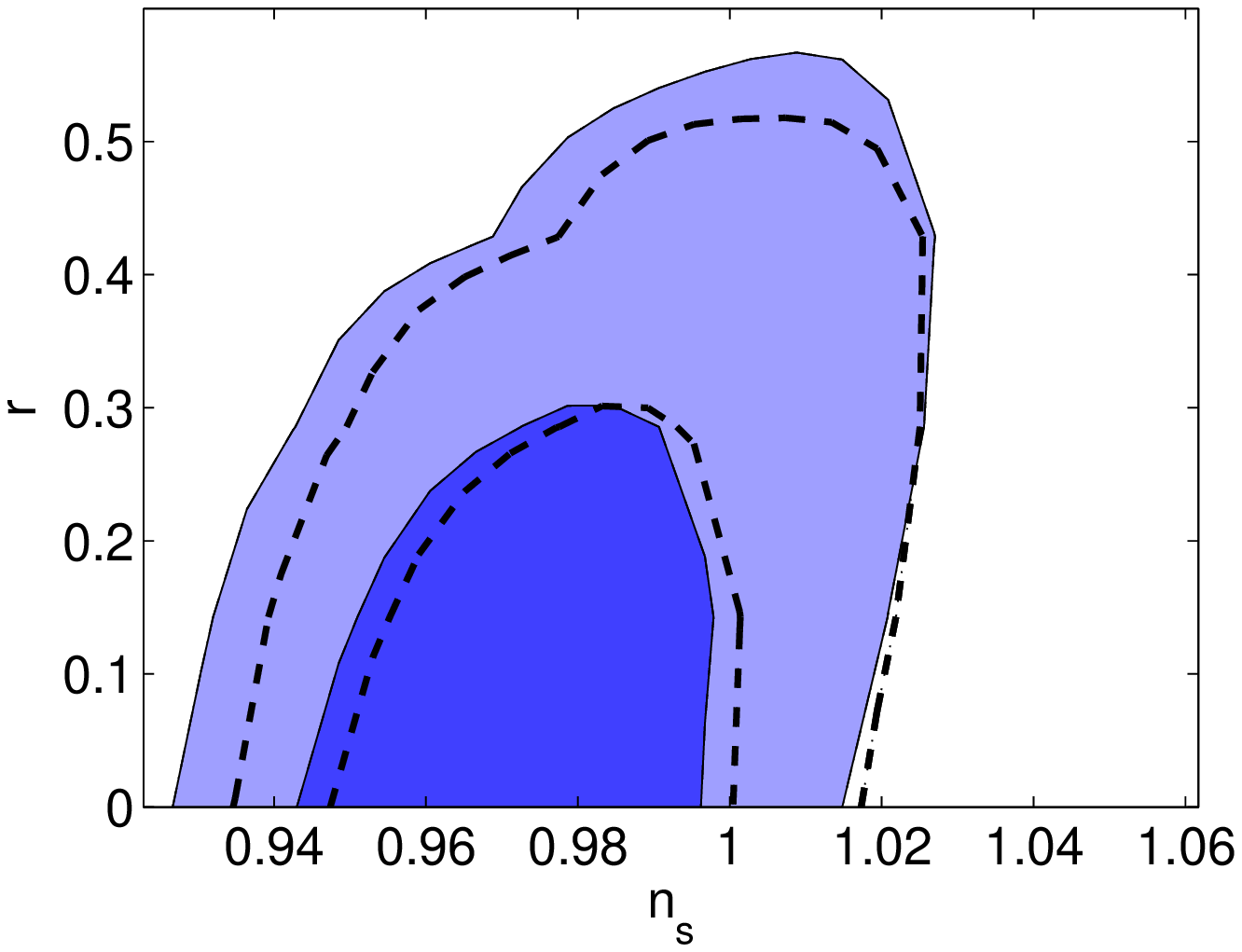}}
\caption{\label{burn-in2} Two-dimensional marginalized distributions
  showing the $68\%$ and $95\%$ confidence level contours for $n_s$
  vs. $dn_s/d\ln k$ (left) and $n_s$ vs. $r$ (right) of interest with
  WMAP 5-year and ACBAR data (empty contours) and and WMAP 5-years
  data+ACBAR data with clustering for point sources (blue contours).}
\end{figure*}

With the addition of ACBAR data, the clustering amplitudes of
unresolved point sources in both WMAP and ACBAR are better
constrained. Though we only use WMAP point source clustering model for
WMAP data only and a separate model for ACBAR point sources at $\ell >
900$, both parameters are better constrained because the combination
of WMAP and ACBAR data pin down the overall cosmological model leaving
less room for the point source piece to change in amplitude.  In
combination, with clustering of point sources included for both WMAP
and ACBAR, we find $n_s=0.959 \pm 0.014$, which is different from the
WMAP+ACBAR value of $n_s=0.963 \pm 0.014$ by about 0.3$\sigma$.

In Fig.~3 we summarize likelihoods of the parameters involved.  As
shown there, when clustering is included the large differences on
cosmological parameter estimates with WMAP data alone appear in
$n_s,\Omega_bh^2$ and $\Omega_ch^2$. However, as discussed for $n_s$,
once we include ACBAR data and with clustering of point sources both
for WMAP and ACBAR, the probability distributions are more consistent
with the WMAP data alone, but with clustering ignored. While it seems
like large multipole data from an experiment such as ACBAR do not
improve cosmological parameters estimated from WMAP, in our case, we
do see an improvement by constraining the point source clustering
amplitude better.  The associated contour plots for parameters that
are mostly affected by clustering of point sources are summarized in
Fig.~4 for combinations of $n_s$ vs. $\Omega_bh^2$ (left) and $n_s$
vs. $\sigma_8$ (right).

\begin{figure}[htbp]
\begin{center}
\includegraphics[scale=0.8,angle=-90,width=8cm]{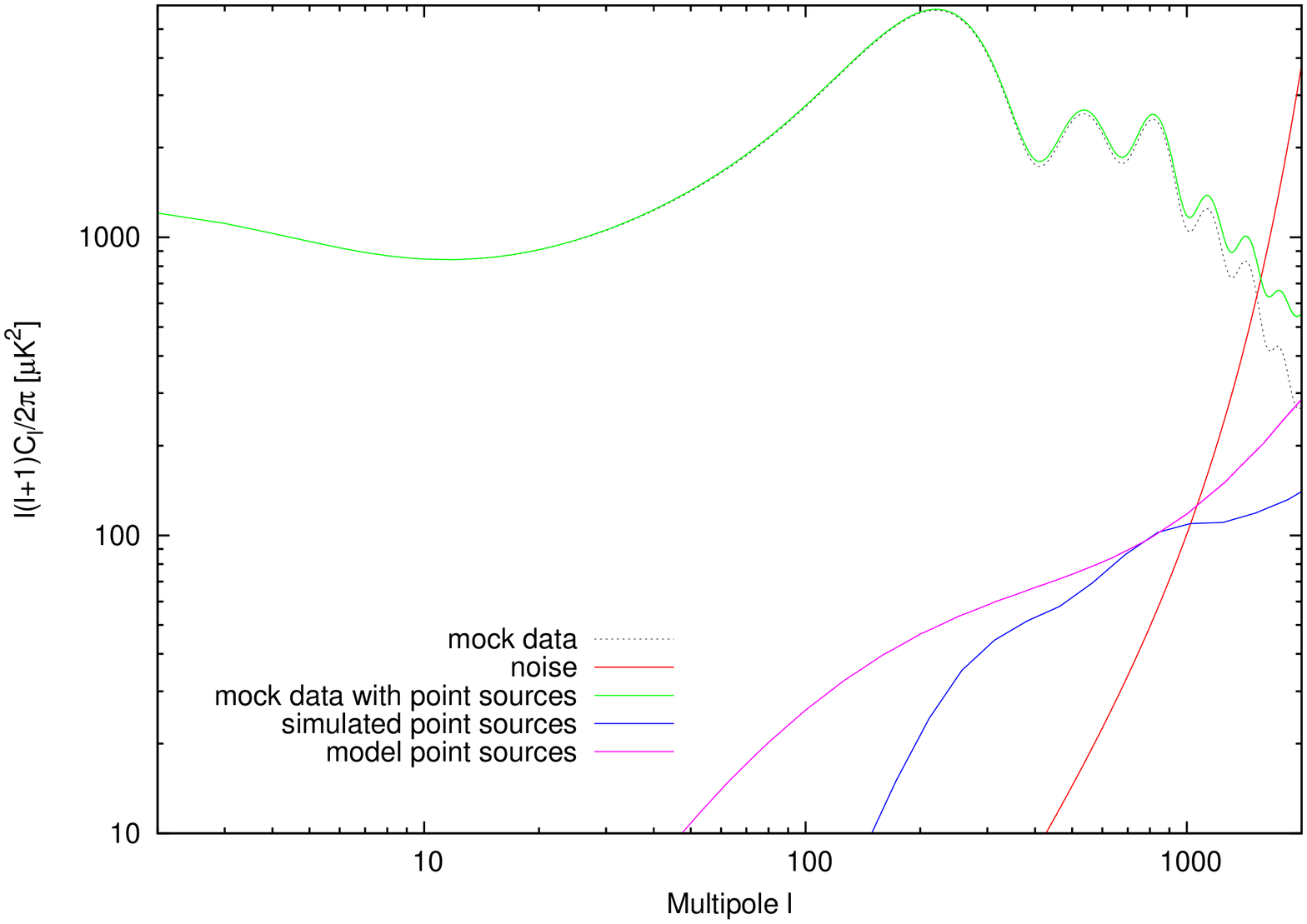}
\end{center}
\caption{Planck mock data with and without a contribution from point
  sources: an offset in the temperature power spectrum is visible for
  small scales. Also showed are the model for point sources used in
  the paper and a simulated signal for point sources from public point source maps
for Planck HFI channels (in this case at 143 GHz) from the Planck team (see text for details).}
\end{figure}

In addition to standard $\Lambda$CDM runs with a power-law power
spectrum for density perturbations, we also study the impact of point
sources on the running of the spectral index and on the
tensor-to-scalar ratio, in addition to the main parameters of the
$\Lambda$CDM cosmological model outlined in Table~I.  In Table~II we
summarize our results. Our results for the combination of WMAP and
ACBAR without clustering are generally consistent with previous
results \cite{Komatsu5yr}, but with minor differences such as a
2$\sigma$ upper limit on $r$ of 0.36 instead of 0.4. The differences
between with and without clustering are also minor and this is
primarily due to the fact that we take the combination of WMAP and
ACBAR.  In Fig.~5 we summarize these results in contour plots with
$n_s$ vs. running (dn$_s/d\ln k$, left) and $n_s$ vs. $r$ (right).

\begin{figure*}[!]
\centering
\subfigure[]{\includegraphics[width=8cm]{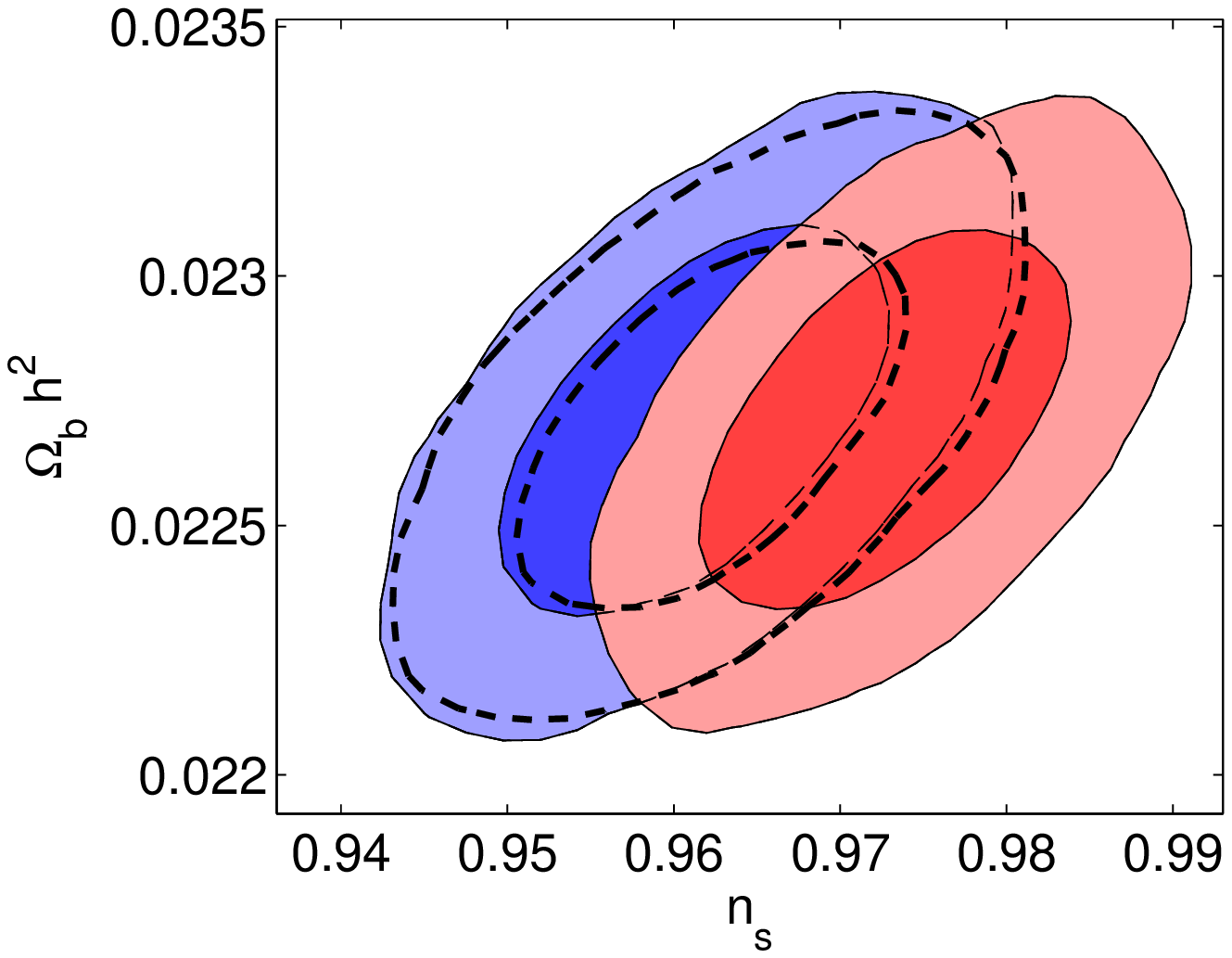}}
\hspace{0.1cm}
\subfigure[]{\includegraphics[width=8cm]{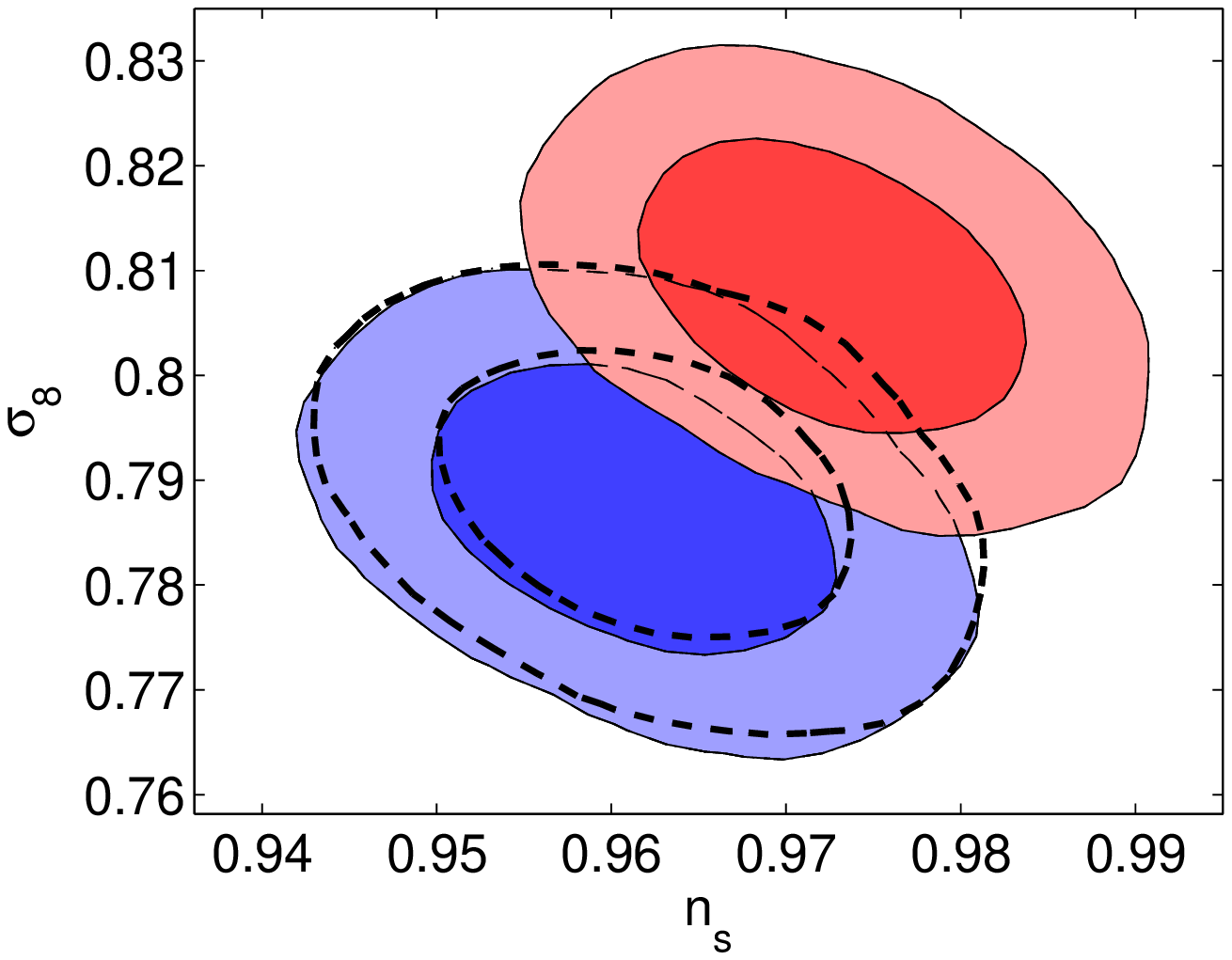}}
\caption{\label{planck2} Two-dimensional marginalized distributions
  showing the $68\%$ and $95\%$ confidence level contours for $n_s$
  vs. $\Omega_bh^2$ (left) and $n_s$ vs. $\sigma_8$ (right) in three different cases: 
Planck mock data alone
  (empty contours), Planck mock data with clustering for point sources 
 marginalized as a shot-noise (red contours, see text for details), and finally
  Planck mock data with a clustered point source signal which is marginalized
  over with a model for clustering. The marginalization over point
  source clustering in Planck completely removes the bias introduced
  by the point sources signal and contours overlap.}
\end{figure*}


\begin{table*}
\caption{Mean values and marginalized $68 \%$ c.l. limits for several
  cosmological parameters from Planck mock data with and without a
  point source contribution.}
\begin{center}
\begin{tabular}{|c|c|c|c|c|c|}
\hline
Parameter & Planck mock data & Planck mock data    & Planck mock data  & Planck mock data    \\ 
          &                  &  with clustering    & with clustering &  with clustering  \\
          &                  & Point sources ignored &  $C_l^c \times (0<P_{Planck}<1)$& Shot-noise only ($C_l^{sn} = P_{\rm Planck}$)\\
\hline \hline
$\Omega_bh^2$&$0.02270_{-0.00024}^{+0.00025}$&$0.02753\pm0.00028$&$0.02271\pm0.00025$& $0.00271_{-0.00024}^{+0.00025}$\\
$\Omega_{\rm c}h^2$&$0.1082\pm0.0019$&$0.0924\pm0.0017$&$0.1080_{-0.0020}^{+0.0019}$ & $0.1092_{-0.0019}^{+0.0020}$   \\
$\Omega_{\Lambda}$&$0.750\pm0.010$&$0.831\pm0.007$& $0.751_{-0.010}^{+0.011}$& $0.745\pm0.010$  \\
$n_s$&$0.962_{-0.007}^{+0.008}$&$1.159\pm0.007$& $0.961_{-0.007}^{+0.008}$& $0.973\pm0.007$\\
$\tau$&$0.090\pm0.007$&$0.260_{-0.016}^{+0.015}$&$0.090_{-0.006}^{+0.007}$&  $0.094_{-0.008}^{+0.007}$\\
$\Delta^2_R$&$(2.41\pm0.10)\cdot10^{-9}$&$(1.85\pm0.10)\cdot10^{-9}$&$(2.41\pm0.10)\cdot10^{-9}$ & $(2.40\pm0.10)\cdot10^{-9}$\\
\hline
$\sigma_8$&$0.789_{-0.009}^{+0.008}$&$0.903_{-0.014}^{+0.015}$&$0.787\pm0.009$& $0.808\pm0.009$ \\
$\Omega_m$&$0.250\pm0.010$&$0.169\pm0.007$& $0.249_{-0.011}^{+0.010}$&  $0.255\pm0.010$     \\
$H_0$&$72.4_{-0.9}^{+1.0}$&$84.3_{-1.2}^{+1.1}$&$72.5\pm1.0$& $72.0\pm1.0$  \\
$z_{reion}$&$11.3\pm0.6$&$19.9_{-0.7}^{+0.8}$&$11.3\pm0.6$& $11.6\pm0.6$\\
$t_0$&$13.69\pm0.04$&$13.04_{-0.04}^{+0.05}$& $13.69_{-0.05}^{+0.04}$& $13.70\pm0.04$\\
$A_{SZ}$&$1.40\pm0.07$&$<2.00$& $1.66_{-0.21}^{+0.23}$& $1.84_{-0.14}^{+0.16}$\\
$P_{\rm Planck}$&$--$&$--$& $<1.00 (2\sigma)$& $C_l^{\rm sn}<3.6\times10^{-5}$ $\mu$k$^2$-sr $(2\sigma)$\\
\hline
\end{tabular}
\label{table:2}
\end{center}
\end{table*}

While we find differences at the level of $1\sigma$ for WMAP data
alone with clustering of point sources, our results show that the
differences are smaller and insignificant once WMAP data are combined
with ACBAR data and using two separate estimates for point source
clustering in WMAP and ACBAR data. In future, Planck data will observe
CMB anisotropies down to smaller angular scales and extending to
higher frequencies where clustering of sources becomes increasingly
important \cite{Scott,GonzalezNuevo:2004fj}. In this case, it is clear
that a simplified approach with a shot-noise for unresolved point
sources in Planck data may not be appropriate when extracting
cosmological parameters.

\subsection{Planck mock data}

To understand how clustering of point sources impact cosmological
parameter determination, we created several mock datasets with noise
properties consistent with Planck 143 GHz channel of HFI and assuming
the best-fit WMAP5 parameters \cite{Komatsu5yr} for cosmology. We
model the point source clustering and the shot-noise by making use of
existing high-frequency data as we did for ACBAR. While we only consider  a
single clustering spectrum, at high-frequencies of Planck HFI, two separate populations
of point sources are expected: radio, dominating at low-frequencies, and sub-mm or far-IR sources
at high-frequencies \cite{Amblard}. Here,  as we only have total number counts at  150 GHz, without
any information on hwo to separate the counts to the two populations, we make use of a single
clustering spectrum. It will be necessary to return  to this topic later once Planck data become available
with additional information, from Planck and Herschel, on the far-IR population in HFI channels.

We summarize our results  related to Planck data in Fig.~6.  In addition to the analytical model of point
source clustering used in this paper based on the halo model, we also made use of publicly available Planck source
maps\footnote{http://www.planck.fr/article334.html} from the Planck
Working Sub-Group for Compact Source Fields to measure the angular
power spectrum of points sources at the Planck HFI 143 GHz channel.
These maps are derived from a model based on GalICS
  model\footnote{http://galics.iap.fr/} using the Mock Map Facility
  (MoMaF,\cite{blaizotetal05}). The power spectrum is computed after
  removing point sources with a flux greater than 72 mJy, 5$\sigma$
  detection level of Planck at that frequency according to
  \cite{fernandez08}.

As shown in Fig.~6, while the amplitude of our analytical model
matches with the residual clustering spectrum of point sources in the
Planck simulation at multipoles of $10^3$, our analytical model has
residual point sources that are more clustered than the simulated
sources. We believe this is due to the finite size of boxes used to
simulate the source distribution by the Planck team.  While we fix our
point sources to the analytical model as shown in Fig.~6, we again
capture the uncertainty in the amplitude with a parameter $P_{\rm
  Planck}$ (in this case clustering and shot-noise combined as in the
case of ACBAR data) and marginalize over this parameter when
estimating cosmological parameters.

We follow the same procedure as fitting existing WMAP and ACBAR data
to extract cosmological parameters with Planck. We use data at $l < 2000$, though Planck analysis can be extended to
higher multipoles, which are likely to be contaminated by additional secondary anisotropies beyond SZ \cite{Cooray5}, and
complications of the non-Gaussian covariance \cite{Cooray6}. The best-fit
cosmological parameters and 68\% confidence errors for the standard
6-parameter $\Lambda$CDM case complemented by $A_{\rm SZ}$ and $P_{\rm
  Planck}$ are tabulated in Table~III. Without point sources, we
recover the best-fit cosmology that was used to create the mock. Once
the mock includes point sources and we ignore the effect of point
sources when model fitting the data, we find that the parameters are
significantly biased; in some parameters this bias is more than
20$\sigma$.   As in the case of current data, we consider the possibility that if it is adequate
to model point sources with just a shot-noise power spectrum. We allow $C_l^{sn} = P_{\rm Planck} \times 0.0075$ $\mu$k$^2$-sr
and fit the data by varying $P_{\rm Planck}$. The values of cosmological parameters with the shot-noise marginalized over
is tabulated in Table~III, in addition to the 2$\sigma$ upper limit on $C_l^{sn}$ from the data. 
With a shot-noise description only in the model fit, 
we find biases in cosmological parameters at the level of 1.5$\sigma$, for example, in the case of the spectral index.

Once we include a model for point source clustering, in addition to the shot-noise,
 and marginalize over the overall amplitude with our parameter $P_{\rm
  Planck}$ to capture the overall clustering amplitude, we find that
the biases in best-fit parameters from the values used for the mock
are removed. We show couple of examples for combinations involving the
scalar spectral index $n_s$ in Fig. 7 and Fig. 8 for $\Omega_ch^2$ and
$\sigma_8$, respectively. As shown, once point source clustering is
included in the fit, cosmological parameter biases are
negligible. Ignoring point source clustering, however, impacts the
measurements significantly, though including only a shot-noise for Planck point
sources still results in an appreciable bias. To be completely safe, we
suggest that a reasonable model for point source clustering and
shot-noise be included in the cosmological parameter analysis with
Planck (at each channel used for cosmological measurements) and the
uncertainty in the modeling or predicting the total point source
contribution be marginalized over.

\section{Summary}

The faint radio point sources that are unresolved in cosmic microwave
background (CMB) anisotropy maps are likely to be a biased tracer of
the large-scale structure.  While the shot-noise contribution to the
angular power spectrum of radio point sources has been considered so
far when extracting cosmological parameters with CMB data, we have
shown here that one should also allow for the possibility of source
clustering. This is especially necessary at high frequencies where the
clustering of sources is expected to dominate the shot-noise level of
the angular power spectrum at tens of arcminute angular scales.  As we
find, the differences seen by the WMAP team for V and W-band angular
power spectra do allow point source clustering, though one can wrongly
conclude clustering is unnecessary if lower frequency data are
included.

Here, we have made an estimate of the clustering of unresolved
radio sources in both WMAP and ACBAR by making use of existing counts
at 95 GHz and by making several assumptions on the sources such as the
redshift distribution.  To account for the uncertainty in modeling
the clustering, we included an extra nuisance parameter for
each dataset and have marginalized over this parameter when model
fitting for cosmological parameters.  For the combination of WMAP
5-year data and ACBAR, we find that the spectral index changes from a
mean value of $0.963 \pm 0.014$ without point-source clustering to a
value of $0.959 \pm 0.014$ when the clustering of point sources
are included in model fits, a difference of $0.3\sigma$.  We also
discussed the full parameter set with clustering of radio point
sources and changes to additional parameters such as $dn_s/d\ln k$ and
the tensor-to-scalar ratio $r$.  While we find that the differences
are marginal with and without source clustering in current data, we
have suggested that it is necessary to account for source clustering
with future datasets such as Planck, especially to properly model fit
anisotropies at arcminute angular scales and using high-frequency
data. For Planck, we find that simply including the point sources as a shot-noise
only out to $l$ of 2000 for cosmological parameter estimation results in biases at
the level of 1.5$\sigma$. While we simply model Planck point sources with a single power
spectrum, since at high frequencies both radio and far-IR  sources are expected to contribute,
it may be necessary to return to a proper model of total unresolved source clustering in
Planck in future.

\begin{center}
{\bf Acknowledgments}
\end{center}
This work was supported by NSF CAREER AST-0645427. We thank Mike Nolta
for clarifying our questions related to the point source modeling by
the WMAP team. AC thanks Dipartimento di Fisica and INFN, Universita'
di Roma-La Sapienza and Aspen Center for Physics for hospitality while
this research was completed. AA acknowledges partial support from a McCue fellowship from
the UCI Center for Cosmology.

\end{document}